\documentclass[%
reprint,
amsmath,amssymb,
aps,
pre,
floatfix,
showkeys
]{revtex4-1}

\usepackage{graphicx}

\usepackage{dcolumn}
\usepackage{bm}
\usepackage{stackrel}
\usepackage{color}

\begin{document}

\newcommand{\jlhd}[1]{{\bf $\langle$JL: #1$\rangle$}}

\title{Damping and clustering into crowded environment of catalytic chemical oscillators}

\author{Carlos Echeverria}
  \email{cecheve@ula.ve}
\author{Jos\'e L. Herrera$^*$}
  \email{jdiestra@ictp-saifr.org}
\author{Kay Tucci$^*$}%
  \email{kay@ula.ve}
\affiliation{%
 $^*$CeSiMo, Facultad de Ingenier\'ia, Universidad de Los Andes, M\'erida 5101, Venezuela. \\
 $^\dagger$ICTP South American Institute for Fundamental Research, IFT-UNESP, S\~ao Paulo, SP Brazil 01440-070. \\
 $^\ddag$SUMA, Facultad de Ciencias, Universidad de Los Andes, M\'erida 5101, Venezuela.}
\author{Orlando Alvarez-Llamoza}
\email{llamoza@gmail.com}
\affiliation{%
Grupo de Investigaci\'on de Simulaci\'on, Modelado, An\'alisis y Accesibilidad. Universidad Cat\'olica de Cuenca, Cuenca 010105, Ecuador.}
\author{Miguel Morales}
\email{mmorales@upsin.edu.mx}
\affiliation{%
Unidad Acad\'emica de Ingenier\'ia en Nanotecnolog\'ia, Universidad Polit\'ecnica de Sinaloa, Mazatl\'an, Sinaloa 82199, M\'exico.}

\date{\today}

\begin{abstract}
A system formed by a crowded environment of catalytic obstacles and complex oscillatory chemical reactions is inquired.
The obstacles are static spheres of equal radius, which are placed in a random way.
The chemical reactions are carried out in a fluid following a multiparticle collision scheme where the mass, energy and local momentum are conserved. 
Firstly, 
it is explored how the presence of catalytic obstacles changes the oscillatory dynamics from a limit cycle to a fix point reached after a damping.
The damping is characterized by the decay constant, which grows linearly with volume fraction for low values of the mesoscale collision time and the catalytic reaction constant.
Additionally, it is shown that, although the distribution of obstacles is random, there are regions in the system where the catalytic chemical reactions are favored. 
This entails that in average the radius of gyrations of catalytic chemical reaction does not match with the radius of gyration of obstacles, that is, clusters of reactions emerge on the catalytic obstacles, even when the diffusion is significant. 
\end{abstract}


\maketitle

\section{\label{sec:Introduction}Introduction}

In chemical oscillations, as in other chemical and biological processes, reaction and diffusion are two of the most basic transport mechanisms underlying their description.
Several numerical and experimental studies have been performed in order to understand these transport mechanisms in homogeneous media.
Additionally, it is known that when the environment is crowded with obstacles the transport processes can be significantly modified \cite{kapral2012chemical,echeveria2007diffusion}.
An important system with a crowded environment is a biological cell, where the volume is occupied by structural elements such as microtubules and filaments, various organelles and a variety of other macromolecular species \cite{goodsell1991inside}. For example, within bacterial cells macromolecules account for more than 40\% of their volume \cite{zimmerman1991estimation}. Under these conditions the fluid and the chemical reactions occurring in it may behave differently from how they do in solute solutions \cite{fulton1982crowded,zimmerman1993macromolecular,ellis2001macromolecular}. 
Understanding such changes and differences is of great importance in processes as essential as protein folding \cite{van1999effects,cheung2005molecular,mittal2010dependence,gnutt2015innenrucktitelbild,echeverria2011mesoscopic,echeverria2012molecular}, protein-protein binding \cite{zimmerman1993macromolecular,minton1981excluded}, gene regulation \cite{matsuda2014macromolecular,morelli2011effects,tan2013molecular,mcguffee2010diffusion,roberts2011noise} and enzyme activity \cite{derham2006effect,norris2011true,zhou2008macromolecular,echeverria2014diffusional,echeverria2015enzyme}, among others.

In a catalytic crowded system, the reaction dynamics of reactive particles introduces new features to the reaction-diffusion kinetics. 
In particular, the rate constants and the diffusion coefficient depend  on the fraction of volume that is occupied by reactive particles in nontrivial ways.
There are experiments where phenomena like synchronization, quorum sensing \cite{TTWHS2009,tinsley2010dynamical} and  emergence of chimeras \cite{nkomo2013chimera,nkomo2016chimera} are modulated by the fraction of the volume occupied by reactive obstacles.
The theoretical treatments of this problem requires that the long-range nature of the diffusive coupling among the reactive obstacles to be properly taken into account \cite{felderhof1976concentration,lebenhaft1979diffusion,gopich2002concentration,echeveria2007diffusion}.

In this paper we explore the behavior of complex oscillating chemical reactions in a catalytic crowded environment with hydrodynamic coupling.
We consider a simple model where hundreds of thousands of small particles undergo motion among a random distribution of stationary catalytic spherical obstacles and compute the dependence of the system's attractor and the decay constant of oscillations with respect to the volume fraction of obstacles and the viscosity of the fluid.
Moreover, by means of the radius of gyration of the obstacles on which the catalytic reactions take place, we study the clustering of the chemical reactions, which could be a signature of some emerging properties induced by the crowding.

Due to the intrinsic difficulties to track analytically the evolution of the system for arbitrary values of the volume fraction and viscosity, our results are obtained from simulations adapting the chemical reactions to the Multiparticle Collision technique (MPC) \cite{MK1999,MK2000,Kapral2008,GIKW2008}.
Although our model is simple, it captures some of the features of crowding effects on oscillatory chemical reactions and sets a starting point for the construction of more detailed models in crowded environments.
Also, an emergent clustering phenomenon is observed in which the mean distance between reactions on catalytic obstacles is smaller than it should be, that is, the reactions are closer than expected.

In Section II we give the details of the model, describing how the Selkov oscillatory chemical reaction is immersed into hydrodynamic fluid with obstacles~\cite{echeverria2010autocatalytic} through a multiparticle collision approach \cite{rohlf2008reactive}.
The  results are presented in Section III. 
The conclusions of the study are given in Section IV.

\section{\label{sec:Model}The model}

\subsection{Reactive multiparticle collision dynamics}

Unlike most studies in this area, which focus on the reactive catalytic event on the obstacles when the particles simply diffuse between them, here we study a situation in which, in addition to the reactive events on the catalyst surfaces, there are complex reactions in the fluid.

The simulations were carried out on a three-dimensional cubic system with volume $V$ that contains a large number of particles, undergoing a reactive dynamics in a field of catalytic obstacles. 
More specifically, the system contains $N = \sum_{\ell=1}^s N_\ell$ point particles with mass $m$, where the sub-index $\ell$ indicates to which of the $s$ species the $N_\ell$ particles belong.

\begin{figure}[hbt]
\includegraphics[width=0.85\linewidth]{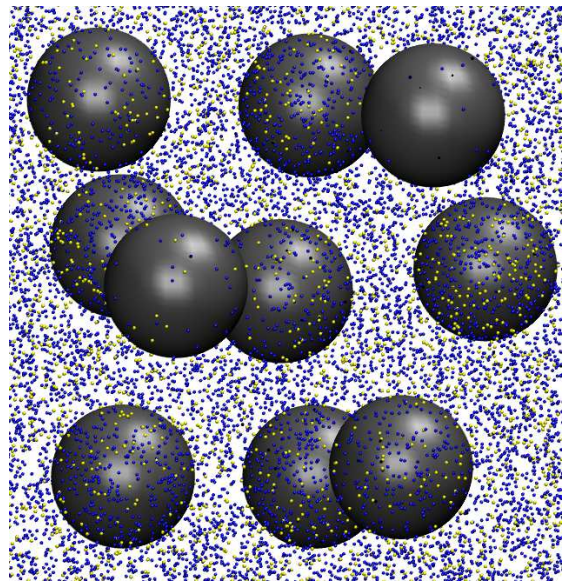}
\caption{\label{fig:system} Snapshot of a typical configuration of the system. The big gray spheres are the catalytic obstacles and the small balls represent the reactive particles $X$ (yellow) and $Y$ (blue) of the fluid.}
\end{figure}

Additionally, the volume contains $N_O$ non-overlapping, identical and immobile catalytic spheric obstacles of radio $\sigma$. 
The volume fraction occupied by the obstacles is $\phi~=~N_O V_O/{V}$,  where $V_O=4 \pi \sigma^3/3$ is the volume of each sphere, being $V_f = {V}(1-\phi)$ the remaining volume free of obstacles.
FIG.~\ref{fig:system} shows a typical configuration of the system, where it can be appreciated the catalytic obstacles and the reactive particles of the fluid.
Periodic boundary conditions were employed in the $N$ point particles displacement, while the $N_O$ obstacles are completely inside the simulation box.

To adapt the chemical reactions to Multiparticle Collision (MPC) \cite{MK1999,MK2000,Kapral2008,GIKW2008}, the simulation time unit is set as the time between collisions where reactive events could occur \cite{RFK2008,KT2005}. 
In MPC the particles have continuous positions and velocities with free stream between multiparticle collision events that occur at discrete times $\tau$. 
To carry out collisions, the volume $V$ is partitioned into $\cal N$ cubic cells of volume $\cal V$, where $V=\cal N \times \cal V$. 
Each cell is labeled with a index $\xi$. 
There are $N_\ell^\xi$ particles of species $\ell$ in cell $\xi$, and the total number of particles in that cell is $N_\xi = \sum_\ell N_\ell^\xi$. 
With a single species ($\ell=1$) the multiparticles collisions are carried out as follows: 
at every time step $\tau$,  a random rotational operator $\hat{\omega}_\xi$ is assigned to each cell. 
The velocity of the center of mass in the cell $\xi$ is ${\bf V}_\xi = N_\xi^{-1} \sum_{i=1}^{N_\xi} {\bf v}_i$, where ${\bf v}_i$ is the velocity of the particle $i$ before the collision. After the collision is performed, the velocity of particle $i$ will be given by ${\bf v}_i' = {\bf V}_\xi + \hat{\omega}_\xi({\bf v}_i + {\bf V}_\xi)$. 

The step-collision rule can be generalized to multicomponent species in the system \cite{TK2004}. If $\ell_i \in \{1,2, \dots, s\}$ denotes the species label of particle $i$, then we may write collision rule as
\begin{equation}
 \label{eq:rule_multicollision}
 {\bf v}_i' = {\bf V}_\xi + \hat{\omega}_\xi({\bf V}_\xi^{\ell_i} - {\bf V}_\xi) +
 \hat{\omega}_\xi^{\ell_i}\hat{\omega}_\xi({\bf v}_i - {\bf V}_\xi^{\ell_i}) \;,
\end{equation}
where ${\bf V}_\xi^{\ell_i}$ is the velocity of the center of mass of particles of species $\ell_i$ in cell $\xi$ and $\hat{\omega}_\xi^{\ell_i}$ is the rotational operator that only acts on the corresponding subset of particles $\ell_i$ in $\xi$.
Both rotational operators, $\hat{\omega}_\xi$ and $\hat{\omega}_\xi^{\ell_i}$, are randomly chosen at each collision step.
This collision rule conserves mass, momentum and energy, and preserves phase space volumes.

In the bulk solutions, we assume that molecules may also undergo chemical reactions of the form
\begin{equation}
 R_\mu : \sum_{\ell = 1}^s \nu_\ell^\mu X_\ell  \stackrel{k_\mu}{\longrightarrow}
 \sum_{\ell = 1}^s \bar{\nu}_\ell^\mu X_\ell \;,
\end{equation}
where $\nu_\ell^\mu$ and $\bar{\nu}_\ell^\mu$ are the stoichiometric coefficients for reaction $R_\mu$, $X_\ell$ is the density of chemical species $\ell$ and $k_\mu$ is the velocity constant of the reaction.

In reactive multiparticle collision dynamics, reactive collisions occur at discrete time intervals $\tau_R$ which is multiple of the time step $\tau$. 
The probability that the reaction $R_\mu$ occurs before any other event in the cell $\xi$ in the interval $\tau_R$ is given by
\begin{equation}
 P_\mu^\xi ({\bf N}^\xi, \tau_R) = \frac{a_\mu^\xi}{a^\xi} (1 - e^{-a^\xi \tau_R})\;,
\end{equation}
where ${\bf N}^\xi$ is the vector of species populations in the cell and $a^\xi = \sum_\mu a_\mu^\xi$.

When there are reactive collisions the probability that a reaction $R_\mu$ will occur in a cell $\xi$ with a free volume ${\cal V}_f^\xi$  during the interval $(t,t+dt)$ is given by \cite{RFK2008}
\begin{equation}
 a_\mu^\xi =  k_\mu ({\cal V}_f^{\xi})h_\mu^\xi \;,
\end{equation}
where the notation $k_\mu({\cal V}_f^\xi)$ indicates that the rate constants have been scaled to take into account the free volume of the cell ${\cal V}_f^\xi$, and $h_\mu^\xi$ is a combinatorial factor that accounts for the number of different ways the reaction can occur in the cell, given by
\begin{equation}
h_\mu^\xi = \prod_{\ell = 1}^s \frac{N_\ell^\xi !}{(N_\ell^\xi - \nu_\ell^\mu)!} \;.
\end{equation}

In the presence of catalytic obstacles, emerges a set of chemical reactions of the form
\begin{equation}
 R_\mu: X_\ell + C \stackrel{k_\mu}{\longrightarrow} X_{\ell'} + C,
\end{equation}
that take place on the surface of the obstacles and converts species $\ell$ into $\ell'$. In such reactions the rate constant is given by
\begin{equation}
k_\mu = P_R \left(\frac{8 \pi k_B T}{m}\right)^{1/2} \frac{\phi}{V_O}\;,
\end{equation}
where $P_R$ is the reaction probability, and the other terms are related to the cross section \cite{TK2004}.
Note that when the volume fraction occupied by catalytic obstacles is large, effects that modify the mass-action chemical rate laws and rate constants emerge. 

\subsection{Selkov reaction with catalytic obstacles}

Specifically, we consider a system comprising a solution of reactive species $\ell = \{A,B,X,Y\}$ that follow the reversible version of the Selkov reaction \cite{selokov1968self}
\begin{eqnarray}
 \label{eq:Rselkov}
 \nonumber
 R_1,R_2:\hspace{1.0 cm} & A &\stackbin[k_{-1}]{k_1}{\rightleftarrows} \hspace{0.3 cm} X \;, \\
 R_3,R_4:\hspace{1.0 cm} & X + 2\,Y &\stackbin[k_{-2}]{k_2}{\rightleftarrows} \hspace{0.2 cm} 3 \,X \;, \\
 \nonumber
 R_5,R_6:\hspace{1.0 cm} & Y &\stackbin[k_{-3}]{k_3}{\rightleftarrows} \hspace{0.3 cm} B \;,
\end{eqnarray}
where $A$ and $B$ denote species with constant concentrations that act as feeds which maintain the system out of equilibrium. 
This is a very simplified model of the phosphofructokinase reaction scheme  portion of the glycolytic cycle that contributes to the oscillations seen in this system \cite{selokov1968self}. 
Selkov reaction is a convenient test case for our study because it is a complex chemical reaction that is real. Also, it is very simple, non-linear, with limit cycles and whose mean field reaction dynamics of the reversible version shows oscillatory and steady-state behaviors and with a well known phase diagram \cite{RRR1981}.

As particles in their free streaming movement can collide with the obstacles and undergo a bounce-back collisions, where particles of the species $X$ change to the species $Y$ with probability $P_C$, in the model, additionally to the reactions in Eq.~(\ref{eq:Rselkov}), we include the following reaction
\begin{equation}\label{eq:Spheres}
 R_7: \hspace{1.0 cm} X +  C \stackbin{k_C}{\longrightarrow} C + Y \;.
\end{equation}
The chemical rate law corresponding to Eqs.~(\ref{eq:Rselkov}-\ref{eq:Spheres}) is given by
\begin{eqnarray}
 \label{ec:sysSelkovXC}
 \nonumber
 \frac{dC_X}{dt} &=& k_1 - k_{-1}C_X - k_{2}C_X C_Y^2 + k_{-2}C_Y^3\\
 & &- k_C C_X C_O, \\
 \label{ec:sysSelkovYC}
 \nonumber
 \frac{dC_Y}{dt} &=& k_2C_X C_Y^2 - k_{-2}C_Y^3 - k_3C_Y + k_{-3} \\
 & &+ k_C C_X C_O, 
\end{eqnarray}
where the constant concentrations of the feed species $A$ and $B$ have been incorporated in the $k_1$ and $k_{-3}$ rate constants.

\section{\label{sec:Results}Results}

Simulations are  performed using the multiparticle collision (MPC) approach. We have set the volume of cells, ${\cal V} = 1$, the rotational operations ${\hat \omega}_\xi$ are taken from the set $\{\pm \pi/2\}$ about randomly chosen axes, the mass of particles of the reactive species $\{A,B,X,Y\}$ is $m=1$, and the radius of the catalytic spheres is $\sigma=2.5$. 

For different values of the fraction of volume occupied by the obstacles $\phi$, the system volume $V$ is adjusted to keep the particles density constant in the remaining volume free of obstacles, $n = N/V_f = 11$, varying the number of particles $N$ in the system as little as possible. The initial concentrations of $X$ and $Y$ are $C_X=3.0$ and $C_Y=0.8$ for all cases.

The temperature in reduced units ($m$,${\cal V}$,$\tau$) is set as $T = 5/12$; hence, particles move a fraction of the length of the cell on average, which introduces the impossibility to maintain Galilean invariance \cite{IK2001}. This is corrected applying a shifting to this invariance. 
The rate constants in the Selkov reaction, which yield to oscillatory dynamics, are $k_1 = 0.0009485$, $k_{-1} = 0.0001$, $k_2 = 0.0004$, $k_{-2} = 0.0004$, $k_3 = 0.001$, and $k_{-3} = 0.0001265$. 

The rate constant $k_C$ that characterizes the reaction on the catalytic sphere can be written as \cite{Kapral1981} $k_C~=~p_C k_m k_D /(k_m + k_D)$, where $k_m$ is an intrinsic rate constant and $k_D = 4\pi D\sigma$ is the Smoluchowski diffusion rate constant where, $D$ is the diffusion coefficient that can be computed  for multiparticle collision dynamics \cite{Kapral2008,GIKW2009} as $D = D_0 (1-\phi)/(1+\phi/2)$, in a first approximation, where $D_0 \approx 0.479\,\tau$ is the diffusion coefficient in a system without obstacles.
The intrinsic rate constant, computed approximately from collision theory, is $k_m = \sigma^2 \sqrt{8\pi k_B T/m}$. 
Table \ref{tab:rates} shows these reaction rates for two different values of $\tau$. These values of the rates mean that as the time elapses the reaction will be increasingly controlled by the diffusion mechanism. 
\begin{table}[htb]
\caption{Approximate values of the intrinsic reaction rate $ k_m $, the Smoluchowski diffusion rate , $k_D$, and the catalytic reaction rate, $k_C$, for two different values of the simulation step, $\tau$, for the simulation setup.\label{tab:rates}}
\begin{ruledtabular}
\begin{tabular}{cccc}
$\tau$ & $k_m$    & $k_D$    & $k_C$ \\
\hline
$1.0$  & $20.225$ & $15.053$ & $8.630$ \\
$0.5$  & $20.225$ & $7.527$  & $5.485$ \\
\end{tabular}
\end{ruledtabular}
\end{table} 

\subsection{Reactions in bulk solution}

The full reaction-diffusion dynamics in the presence of an arbitrary number of catalytic obstacles can be simulated using reactive multiparticle collision dynamics. As described in \cite{RFK2008}, for long time scales in a well mixed system, this mesoscopic dynamics reduces to the mean-field, mass-action equations of chemical kinetics.

\begin{figure}[hbt]
\includegraphics[scale=0.31]{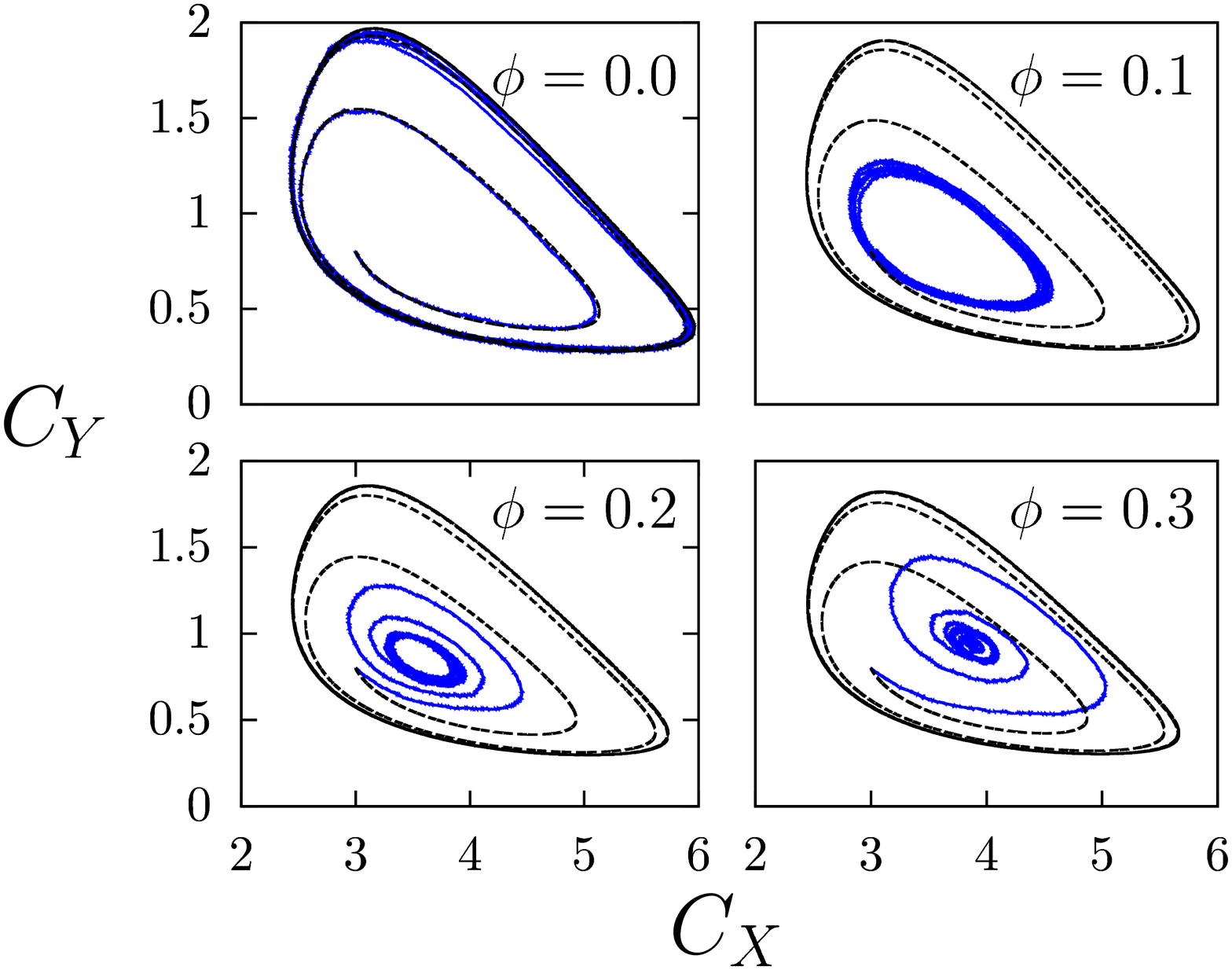} \\
\includegraphics[scale=0.31]{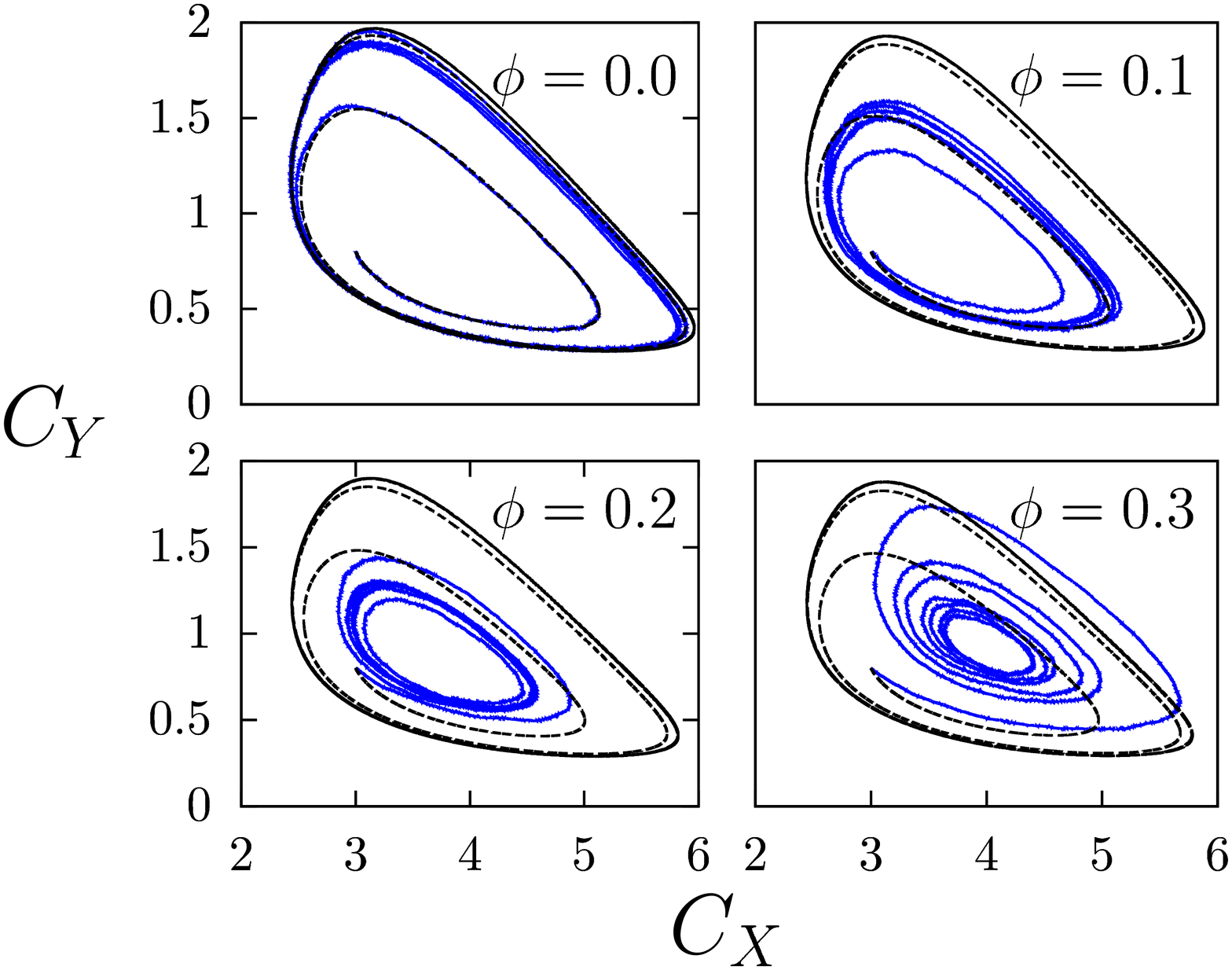}
\caption{\label{fig:phases_space} Trajectories in the phase-space, $(C_X(t), C_Y(t))$,
for $P_C = 4 \times 10^{-5}$ and various values of $\phi$. $\tau = 1.0$ (top) and $\tau = 0.5$ (bottom). The solid blue and black lines are the simulation and mean-field results, respectively. Dashed line represent the attractor of the mean-field model. Simulation time is $t=10^5$ iterations.}
\end{figure}

The simulation results for the globally averaged concentrations $X$ and $Y$ are compared with the mean-field concentrations as trajectories in the phase-space $(C_X(t),C_Y(t))$, as shown in FIG.~\ref{fig:phases_space} for two different values of time step $\tau = 1.0$ (top) and $\tau=0.5$ (bottom). It is noticeable that as $\phi$ increases, the limit cycle in the phase-space reduces until reaching a fixed point. This behavior appears for both values of $\tau$; however, being more discernible for $\tau=1.0$. 
We observe that for $\phi = 0.1$ and $\phi=0.2$ the steady state is a limit cycle, which size depends on $\tau$, being larger for smaller values of $\tau$. Additionally, the limit cycle vanishes for $\phi=0.3$ and $\tau=1.0$, indicating that the diffusion is determinant in the steady state. 

As $\phi$ increases, the mean-field approximation fails to describe the behavior of the system, because the effects of diffusion (in the cyclic limit) are small. To observe the effects of diffusion in this regime we calculate the instantaneous difference of concentration $X$ for two values of $\tau$, $\Delta C_X = C_X^{(\tau=1.0)}(t) - C_X^{(\tau=0.5)}(t)$. 

\begin{figure}[hbt]
\includegraphics[scale=0.31]{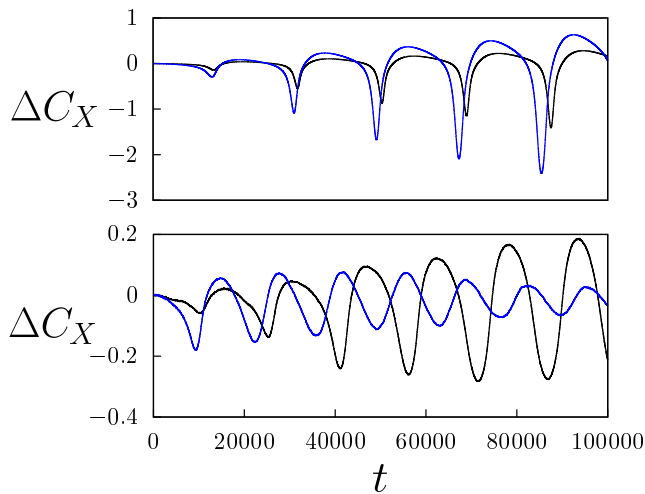} 
\caption{\label{fig:Delta_mass} Evolution of the difference between X concentrations for two different values of the time step, $\tau=1.0$ and $\tau=0.5$. 
The evolution of $\Delta C_X$ is calculated for $\phi=0.1$ (black lines) and $\phi=0.3$ (blue lines) with the same parameters used in FIG.~\ref{fig:phases_space}.
Top: Results obtained using the mean field theory. 
Bottom: Results obtained from the simulations of the MPC model.}
\end{figure}

FIG.~\ref{fig:Delta_mass} shows the evolution of $\Delta C_X$ for two values of $\phi$.
For the mean field approximation (top) $\Delta C_X$ increases with time for both values of $\phi$, taking larger values for larger values of the density $\phi$.
On the other hand, the behavior of $\Delta C_X$  for the simulations (bottom) only increases for small values of density, $\phi=0.1$, while for larger values, $\phi=0.3$, the difference $\Delta C_X$ falls into constant oscillations. We attribute this behavior to the presence of two different steady states in the system, a fixed point and a limit cycle.

\subsection{Behavior of transients}

\begin{figure}[hbt]
\includegraphics[scale=0.31]{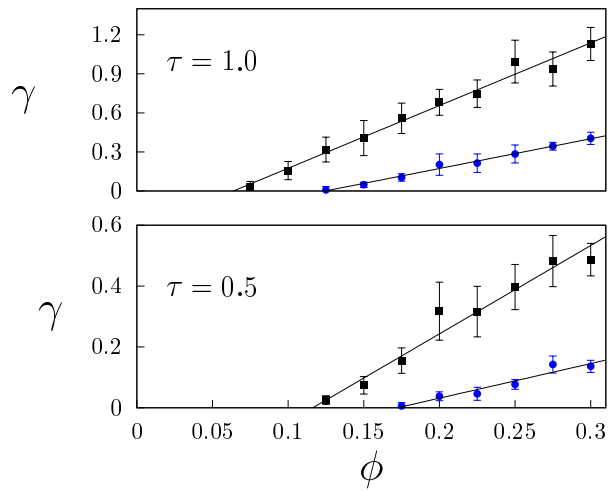} 
\caption{\label{fig:gamma_mass} Decay constant, $\gamma$, as function of catalytic spheres density, $\phi$, for two values of reaction probability, $P_C = 4 \times 10^{-5}$ (circles blue points) and $P_C=1 \times 10^{-4}$ (squares black points); and two values of time step, $\tau = 1.0$ (top) and $\tau=0.5$ (bottom). Error bars are the standard deviations computed over $8$ realizations.}
\end{figure}

Another characteristic of the dynamics that could be appreciated in FIG.~\ref{fig:phases_space} is that for a given value of $\phi$, the amplitude of
oscillations start to converge either to a cycle limit or a fixed point. 
This behavior resembles that of a mass connected to a spring in the presence of a frictional force, and can be described by
\begin{equation}
 \label{ec:osc_fric}
 C_X(t) \sim \cos (\omega t + d) \exp (-\gamma t)\;,
\end{equation}
where the frequency ($\omega$), the phase shift ($d$), and the decay constant $\gamma$ are fitting parameters. In a mechanical system, $\gamma$ represents the friction; however, a more accurate interpretation of $\gamma$ for our system is that it's related with the transport properties of media and the value of $P_C$.
To verify this relationship, FIG.~\ref{fig:gamma_mass} shows how $\gamma$ depends on the density of catalytic spheres $\phi$, that is, on the reaction surface. Note that in all cases the dependency is linear but the slope changes for the different values of $P_C$ and $\tau$, meaning that $\gamma$ effectively depends on the probability $P_C$ and diffusion coefficient in a system without obstacles $D_0$, according to the linear expression $\gamma \sim m_{P_C}^{\tau} \phi$.

In this way, table~\ref{tab:gamma} shows the ratios among the different values of
$m_{P_C}^\tau$ and compare them with the ratios between values of $\tau$ and $P_C$.
\begin{table}[hbt]
\caption{\label{tab:gamma} Ratios between the values of $\tau$, $P_C$ and $m_{P_C}^{\tau}$. To simplify the notation all values of $P_C$ are shown multiplied by $10^5$.}
\begin{ruledtabular}
\begin{tabular}{ccc}
$\tau$ Ratio & $P_C$ Ratio & $m_{P_C}^{\tau}$ Ratio  \\
\hline
 $\displaystyle 1.0/0.5 = 2.0$ & $\displaystyle 10/10 = 1.0$ &
 $\displaystyle \left(m_{10}^{1.0}/m_{10}^{0.5}\right) \approx 1.66$ \\
 $\displaystyle 1.0/0.5 = 2.0$ & $\displaystyle 04/04 = 1.0$ &
 $\displaystyle \left(m_{04}^{1.0}/m_{04}^{0.5}\right) \approx 2.01$ \\
 $\displaystyle 1.0/1.0 = 1.0$ & $\displaystyle 10/04 = 2.5$ &  
 $\displaystyle \left(m_{10}^{1.0}/m_{04}^{1.0}\right) \approx 2.10$ \\
 $\displaystyle 0.5/0.5 = 1.0$ & $\displaystyle 10/04 = 2.5$ &
 $\displaystyle \left(m_{10}^{0.5}/m_{04}^{0.5}\right) \approx 2.56$
\end{tabular}
\end{ruledtabular}
\end{table}

In the table we observe that the better correlations occur for $P_C = 4 \times 10^{-5}$ when $m^\tau_{P_C}$ ratio is 2.01 and $\tau$ ratio is equal to 2; and for $\tau=0.5$ when $m^\tau_{P_C}$ ratio is 2.56 and $P_C$ ratio is 2.5. 
In both cases, the values of $m_{P_C}^{\tau}$ that are closer to $\tau$ and $P_C$ are smaller than these parameters.
In other words, when the velocities of catalytic reactions and diffusion are slow, there is a linear effect on the damping, while for the opposite case, the effect is smaller than linear.
Although the relationship between $m^\tau_{P_C}$, $\tau$ and $P_C$ is not trivial, we can see that the change in the slope of the damping is linear respect to variations of $\tau$ and $P_C$ when the reaction is sufficiently slow, this is $P_C=4\times 10^5$, or when the diffusion is sufficiently slow, this is $\tau=0.5$.

\begin{figure}[hbt]
\includegraphics[scale=0.31]{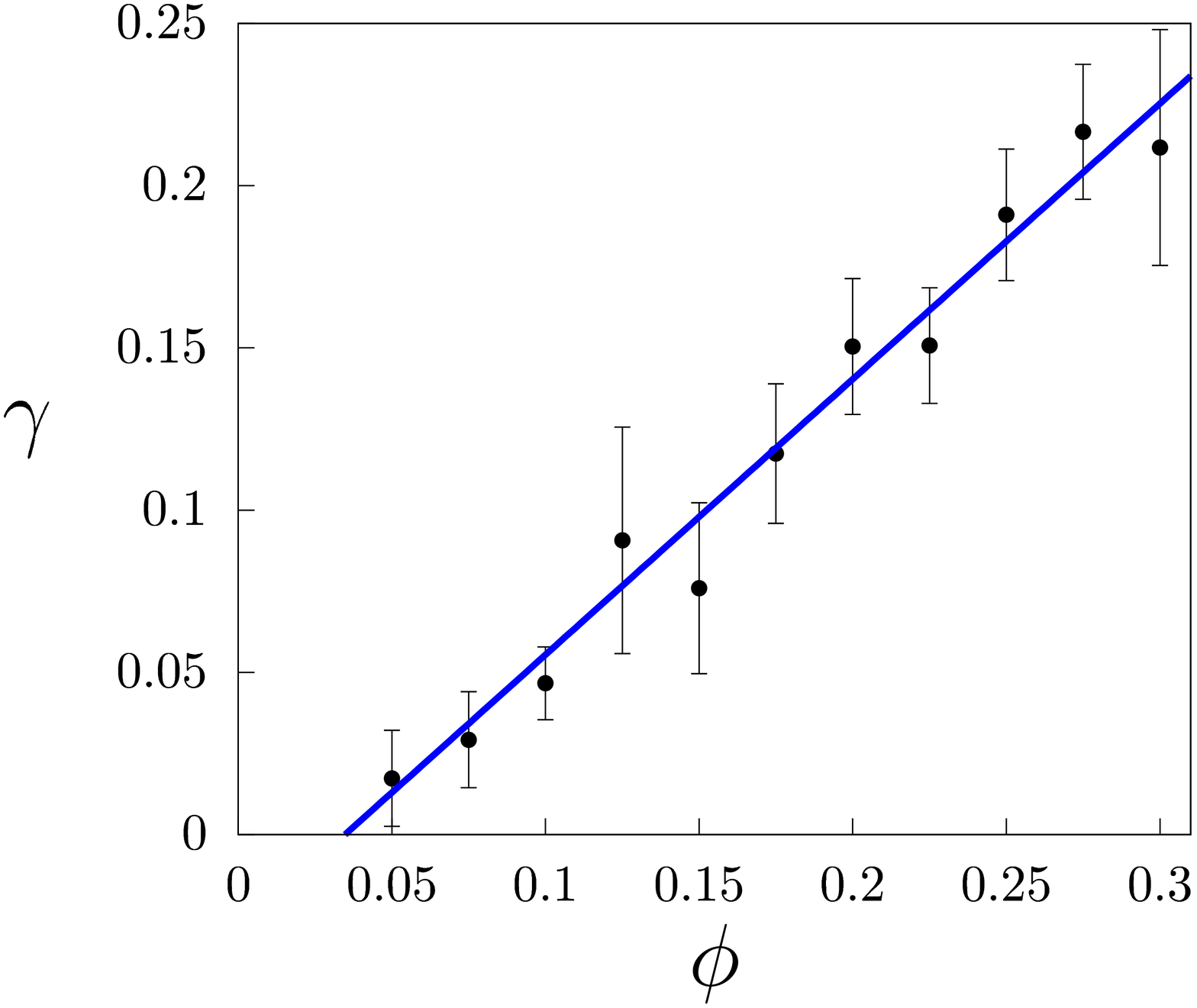} 
\caption{\label{fig:gamma_eV} Decay constant $\gamma$ as function of the volume fraction of obstacles $\phi$ for  reaction probability $\widetilde{P_C}~=~P_C^0 N_O^0/N_O(\phi)$. Points represent the mean values averaged over 8 realization with $P_C^0 = 6 \times 10^{-4}$, $N_O^0 = 10$ and $\tau = 0.5$. Error bars show the standard deviations around the mean.
The line is the best fit of a linear function among of points with a slope approximately of 0.85.}
\end{figure}

We know that when the number of catalytic spheres increases the surface of reaction grows, then the value $k_C$ becomes greater. 
This effect could be suppressed setting the reaction probability on catalytic obstacles  $\widetilde{P_C} = P_C^0 N_C^0/N_O(\phi)$, where $P_C^0$ and $N_O^0$ are the reaction probability and the volume fraction occupied by the catalytic spheres with which the system has the desired reaction rate. 
FIG.~\ref{fig:gamma_eV} shows the decay constant $\gamma$ as a function of the obstacle volume fraction $\phi$ using $\widetilde{P_C}$ as reaction probability on catalytic obstacles. 
Note that despite compensating the increase of the reaction surface with the decrease of $\widetilde{P_C}$, the value of the damping factor changes, increasing linearly with an approximate slope equal to $0.85$.
This shows us that the effect of the spatial distribution of the catalytic spheres is significant on oscillatory chemical reaction.

\subsection{Reactions on catalytic sphere}

Reactions on the surface depend on the concentration of the species $X$ and on its transport properties.
Consequently, if $C_X(t)$ oscillates the number of reactions on the catalytic spheres will oscillate as well.
Despite this oscillations, the cooperative effects of reactions over spheres can be analyzed observing the radius of gyration of the spheres where the reactions take place during each $\tau_R$ interval time given by
\begin{equation}
\rho_g^R = \sqrt{\frac{1}{N_R} \sum_{i=1}^{N_R} \left({\bf r}_i - {\bf r}_P\right)^2}\;,
\end{equation}
where $N_R$ is the number of reaction events that occurred on the catalytic spheres in an interval $\tau_R$, ${\bf r}_i$ is the position of the sphere where the $i$-th event took place, and ${\bf r}_P$ is the center of mass of all catalytic spheres involved in the reaction during the time interval. 
In a random process, reactions can occur on the surface of any of the catalytic obstacles with the same probability and in such case the difference between the radius of gyration of all obstacles ($\rho_g^O$) and the average value over all time intervals of the radii of gyration of obstacles where chemical reactions take place ($\overline{\rho_g^R}$) should be equal to zero, that is $\Delta \rho_g = \rho_g^O - \overline{\rho_g^R} \approx 0$. 
However, if the distribution of obstacles and the diffusion process favor the triggering of reactions on the obstacles in some regions of the volume $V$, there should exist a difference between these two radii of gyration, that is, $\Delta \rho_g \neq 0$.

\begin{figure}[hbt]
\includegraphics[scale=0.3]{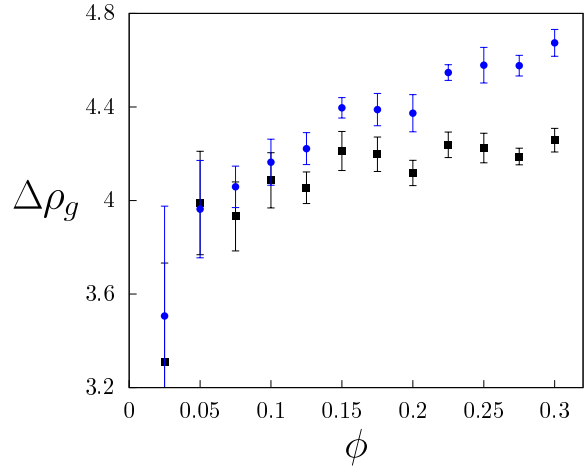}
\caption{\label{fig:delta_Rg} Effect of the volume fraction occupies by the catalytic spheres, $\phi$, on the difference between the radius of gyration of the obstacles and the average values over all time intervals of the radii of gyration of obstacles where chemical reactions take place, $\Delta \rho_g$. Simulations are done with $p_C=10^{-4}$,
$\tau = 1.0$ (black squares) and $\tau = 0.5$ (blue circles).
Error bars shows the standard deviations over $8$ realizations.}
\end{figure}

FIG.~\ref{fig:delta_Rg} evidences that $\Delta \rho_g$ is greater than zero for all values of $\phi$ and for both values of the time step, $\tau~=~1.0$ (black squares) and $\tau~=~0.5$ (blue circles); i.e. in all studied cases $\overline{\rho_g^R}$ is less than expected, that is, reactions on the catalytic spheres are forming clusters.
It can also be appreciated that for low values of the volume fraction of catalytic obstacles ($\phi \leq 0.1$) the values of $\Delta\rho_g$ are independent of the values of $\tau$ considered, which implies that $\overline{\rho_g^R}$ does not depend significantly of the diffusion constant in this range of values of $\phi$. 
Nonetheless, for greater values of $\phi$ the difference $\Delta \rho_g $ only increases when the diffusion is low ($\tau = 0.5$), while for higher diffusion ($\tau = 1.0$) its value reaches an apparent constant behavior. 

As an example of a system where these properties can be determining, there is an experiment published by Taylor et al. \cite{TTWHS2009} where they studied large populations of discrete chemical oscillators that presents synchronized oscillatory behavior. 
The experiment was carried out with $\phi \approx 0.05$ in an agitated medium, in other words, a medium where the diffusion coefficient $D_0$ is large. 
In our model, these parameter values correspond to a system with $\overline{\rho_g^R}$ less than expected and where clustering of the reactions does not depend significantly on the diffusion coefficient. 
The clustering of the catalytic reactions observed in the experiment \cite{TTWHS2009} as well as in our model, suggests that the system's behavior reported by Taylor et al. could be in part due to phenomena that is also present in our model.

\begin{figure*}[hbt]
\includegraphics[scale=0.3]{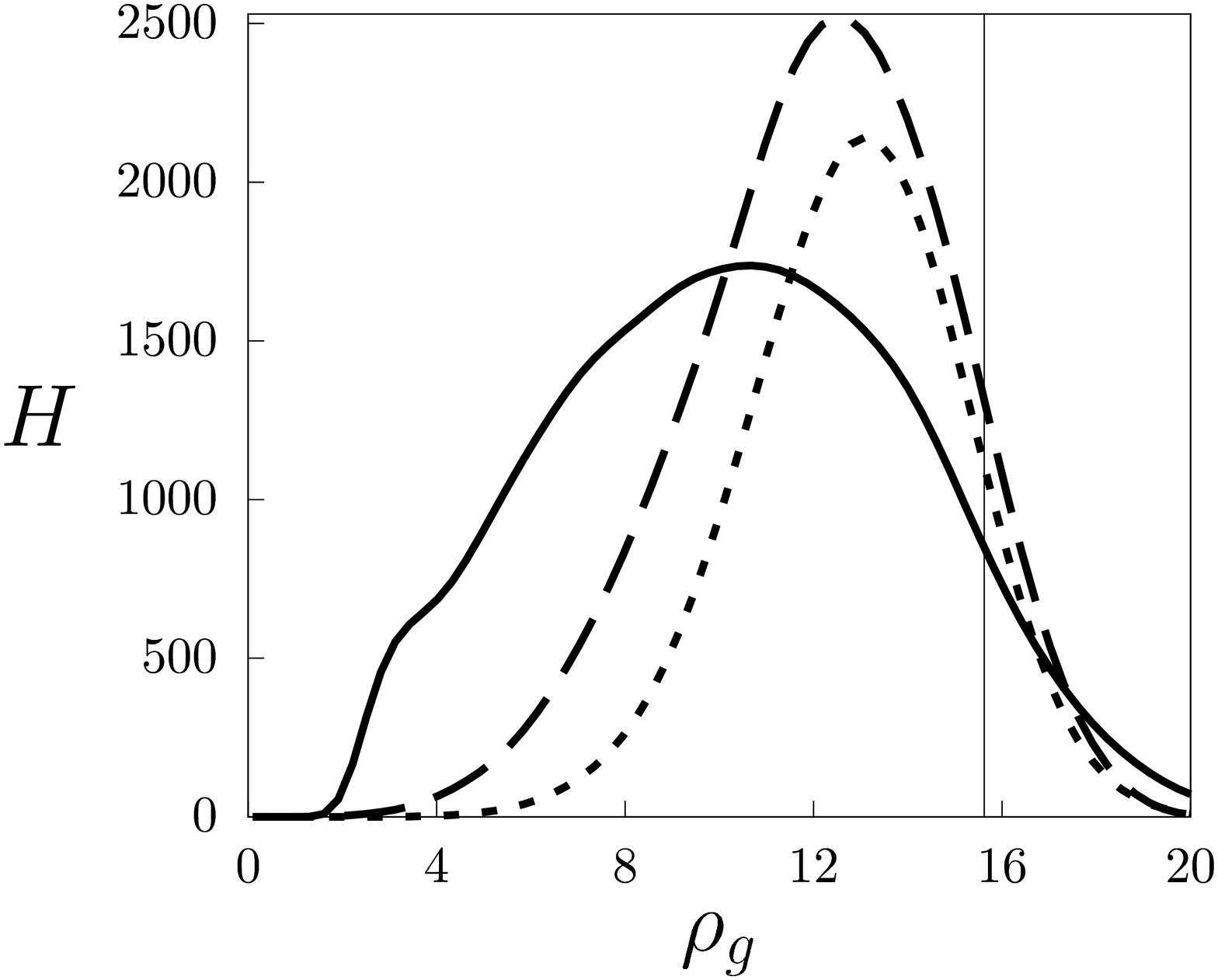}
\includegraphics[scale=0.3]{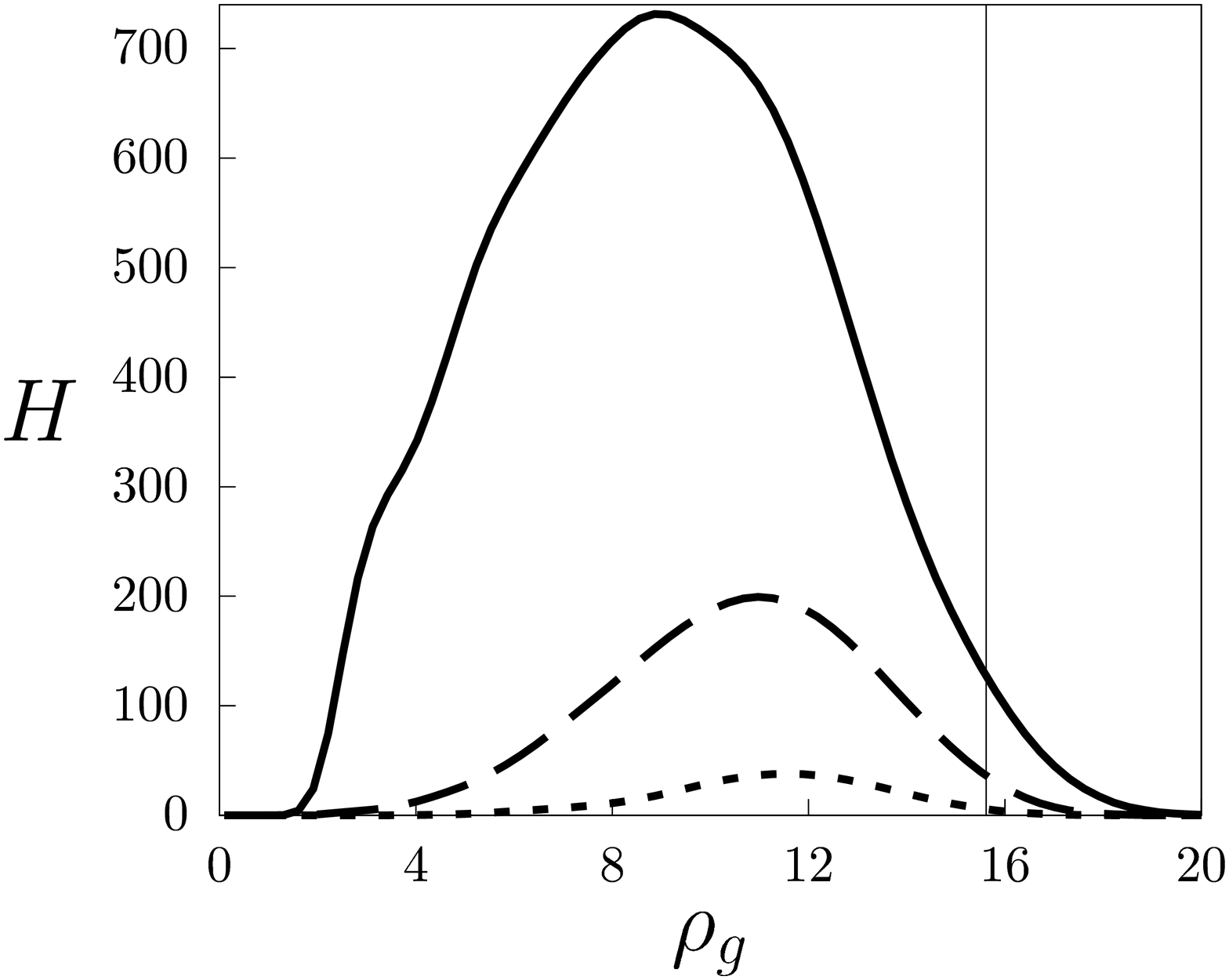}
\caption{\label{fig:Hist_Rg} Average distribution of the the number of reactions that occur during an interval $\tau_R$ with the same radius of gyration of the obstacles where the reactions take place, $n_R$, as a function of their respective radii of gyration, $\rho_g^R$.  
Averages were made over 8 simulations with $\phi = 0.3$, $P_C = 10^{-4}$ and $\tau = 1.0$ (left), and $\tau = 0.5$ (right).
Solid, dashed and dotted lines represent the distributions of $n_R=2, 3$ and $4$ respectively.
The vertical line indicates the mean value of the radius of gyration of all obstacles, $\rho_g^O$.}
\end{figure*}

To understand the nature of the clustering, we calculate the number of reactions that occur during the time interval $\tau_R$, and denote it by $n_R$. 
FIG~\ref{fig:Hist_Rg} shows the average distribution ($H$) of the first three values of $n_R$ as a function of their respective $\rho_g^R$. To create the distributions, we counted, for each $\rho_g^R$,  how many times each $n_R$ occurred during a simulation, averaging these results over 8 realizations.
Note that the results of the simulations, which were carried out in systems with a volume fraction of obstacles $\phi=0.3$ and a probability of reaction $P_C=10^{-4}$, show that most groups of reactions take place in obstacles which radii of gyration are less than the radius of gyration of all obstacles, $\rho_g^O$, indicated in the FIG~\ref{fig:Hist_Rg} with vertical lines. This behavior is observed for both time steps, $\tau=1.0$ (left) and $\tau=0.5$ (right). 

Notice that with greater diffusion, the maximum values of the distributions are closer to the value $\rho_g^O$ and the number of reactions is considerably larger. 
Additionally, it can be observed that the structures formed by the reactions depend on the diffusion mechanism when the closeness among reactions is favored by a greater diffusion rate.
In both cases it is also appreciated, as expected, that as $n_R$ increases, the corresponding average radius of gyration approaches to $\rho_g^O$.
Finally, for sufficiently large values of diffusion, once more we can see that the obstacles, where small groups of catalytic reactions take place, have an average radius of gyration ($\overline{\rho_g^R}$) which is clearly smaller than the radius of gyration of all the obstacles present in the system ($\rho_g^O$), in other words, in the system emerge clusters of catalytic reactions.

\section{Conclusion}

We show that, as previously reported \cite{robertson2015nanomotor}, the effect of the introduction of catalytic obstacles in a system with a complex oscillating chemical reaction is to reduce the oscillations in it, leading the system to a fixed point where the limit cycles are shifted. 
Additionally, we fit these oscillations to a damped periodic function, finding that its decay constant $\gamma$ scales linearly with the  volume fraction occupied by the obstacles ($\phi$) after a certain amount of obstacles is introduced; meaning that there exists a critical damping value for $\phi$. 
Our results show that when the time step ($\tau$) and the probability of reaction on the catalytic spheres ($P_C$) take low values, the slope of the damping ($m_{P_C}^{\tau}$) is linear with respect to them.
Nonetheless, when either the diffusion effects or the rate of catalytic reactions begin to be significant, the crowding of obstacles in the system becomes a crucial factor, even when the reaction probability $p_C^0$ is adjusted to keep constant the quantity of reactions that occur on the catalytic surface of the spheres per time unit.

Furthermore we find that, as a result of the presence of catalytic obstacles in the system, on average the radius of gyration of obstacles where chemical reactions take place is smaller than expected ($\rho_g^R < \overline{\rho_g^O}$); that is, a clustering effect of the catalytic reactions emerges in the system. 
In addition, there is a range of values of the volume fraction occupied by the obstacles ($\phi \leq 0.1$) where the average radius of gyration $\overline{\rho_g^R}$ does not seem to depend significantly on the constant of diffusion $D_0$.
In other words, the model of oscillatory chemical reactions presented in this document is able to show and measure how changes in diffusive transport properties have an important role in the distribution of catalytic reactions in a crowed environment, where hundreds of thousands of particles are involved, which could explain the emergence of phenomena such as the quorum sensing and the chimeric patterns, observed in experiments with this kind of systems~\cite{TTWHS2009,TNS2012}.

\textbf{Acknowledgements}
Jos\'e L. Herrera Diestra is supported by the S\~ao Paulo Research Foundation (FAPESP) under grants 2016/01343-7 and 2017/00344-2

\bibliography{apssamp}

\end{document}